\documentclass{JFM-FLM_Au}
\hypersetup{citecolor=blue,linkcolor=blue}
\usepackage[T1]{fontenc}
\usepackage{amsmath,amssymb}
\usepackage[mathlines]{lineno}

\makeatletter
\renewcommand{\ps@headings}{%
  \let\@mkboth\markboth
  \def\@oddhead{\hfill{\itshape\@righttitle}\hfill}%
  \def\@evenhead{\hfill\itshape\@lefttitle\hfill}%
  \def\@oddfoot{\hfil\thepage\hfil}%
  \def\@evenfoot{\hfil\thepage\hfil}%
  \def\sectionmark##1{\markboth{##1}{}}%
  \def\subsectionmark##1{\markright{##1}}%
}
\makeatother
\newcommand{\ii}{\mathrm{i}}
\newcommand{\sgn}{\operatorname{sgn}}

\lefttitle{P. Simson}
\righttitle{Preprint}

\title{Hamiltonian symmetry behind the vanishing of four-wave interactions at non-trivial resonances for one-dimensional deep-water waves}

\author{P\"aivo Simson\aff{1}}

\affiliation{\aff{1}Laboratory of Solid Mechanics, Department of Cybernetics, Tallinn University of Technology, Tallinn, Estonia
}

\corresau{P\"aivo Simson, \email{paivo.simson@taltech.ee}}

\begin{document}
%\linenumbers

\maketitle

\begin{abstract}
The unexpected cancellation of non-trivial resonant four-wave interactions is the foundation of
several compact and accurate models for one-dimensional deep-water waves. The reasons for this cancellation are not yet fully understood. An additional integral
of motion would account for it, but to our knowledge no corresponding symmetry has been
identified. Here it is shown that in real canonical variables with the cubic terms eliminated,
the fourth-order Hamiltonian is invariant under a sign-dependent phase rotation generated by the
Hilbert transform. The corresponding Noether charge is the action imbalance between right- and
left-moving waves, conserved exactly although number-changing quartic terms remain. The symmetry
confines the quartic kernel to the sector with two wavenumbers of each sign, which contains no
non-trivial resonance, so the cancellation follows without calculation. In complex normal
variables the symmetry is the vanishing of the reduced Zakharov kernel whenever the product of
the four wavenumbers is non-positive, and the known closed form of that kernel follows
directly. The symmetry is the fourth-order instance of a selection rule valid at every order in
these variables: no term of the Hamiltonian has a single wavenumber of the minority sign. The
rule rests on the holomorphic structure of planar deep-water flow and on its square-root dispersion relation.

\end{abstract}


\section{Introduction}\label{sec:intro}
\citet{DyachenkoZakharov1994} computed the four-wave interaction
coefficient of one-dimensional deep-water gravity waves in the reduced
Zakharov formulation \citep{Zakharov1968,Krasitskii1994} and found that
it vanishes identically on the non-trivial resonant manifold. They
remarked that such a cancellation would occur naturally if the system
were integrable, or had at least an additional integral of motion in the
sense of \citet{ZAKHAROV1988283,ZakharovShulman1991}. \citet{CraigWorfolk1995} confirmed the result and showed that the fourth-order Birkhoff normal form for one-dimensional deep-water waves 
is integrable \citep[see also][and references therein]{BertiFeolaPusateri2022}. 
In both cases, computer algebra was used to perform the lengthy calculations.
This unexpected cancellation made it possible to simplify the reduced Zakharov equation greatly, which led to the compact \citep{DyachenkoZakharov2011,DyachenkoZakharov2012,Dyachenko2016_book} and super compact \citep{DyachenkoKachulinZakharov2017envelope,DyachenkoKachulinZakharov2017} equations for unidirectional waves.  These remarkably simple models accurately capture strongly nonlinear phenomena such as modulational instability, breathers, and initial stages of wave breaking, for steepness values up to 0.4.
\citet{DyachenkoKachulinZakharov2017} also stated
a stronger property without derivation (DKZ's sign rule): the four-wave interaction coefficient vanishes
whenever the product of the four wavenumbers is non-positive, on and off the
resonant manifold. Based on this fact, \citet{Dyachenko2019fluids,Dyachenko2020canonical} derived a bi-directional canonical system of super compact equations. He found that
the wave actions of right- and left-moving waves are conserved separately, and
presented these as new integrals of motion. Independently, \citet{Neveu2021} found several conserved nonlocal charges in physical variables by direct perturbative calculation, without identifying their origin or giving a physical interpretation.

Here we show that these seemingly separate facts all reflect one symmetry of the fourth-order Hamiltonian, which we call the Hilbert rotation (\S\ref{sec:symmetry_and_W}). We work in the real canonical variables $(u,v)$ of \citet{Simson2026}, which are Zakharov's $(\eta,\phi)$ after the near-identity canonical
transformation that removes the cubic terms from the Hamiltonian.
The Noether charge of the symmetry, $W=\int vHu\,\mathrm{d}x$, is the action imbalance between right- and left-moving waves (\S\ref{sec:symmetry_and_W}). It is conserved exactly although the number-changing quartic terms are still present in the Hamiltonian, and the total wave action is not conserved. $W$ is the integral of motion mentioned above, and in physical variables it is Neveu's first charge through the order retained. The quartic kernel is elementary in these variables with no frequency denominators, and where it vanishes, on and off the resonant manifold, is decided by the signs of the four wavenumbers alone (\S\ref{sec:kernel_and_signs}). Two derivations become straightforward: the vanishing of \citet{DyachenkoZakharov1994} reduces to a count of wavenumber signs (\S\ref{sec:vanishing_demonstration}), and the simple closed form of the reduced Zakharov kernel \citep{DyachenkoKachulinZakharov2017, Dyachenko2020canonical} is obtained as a projection of the real kernel onto one interaction class, where the symmetry appears as the above-mentioned sign rule (\S\ref{sec:zakharov_section}).

Finally, we trace the symmetry to its origin (\S\ref{sec:origin}). In the
frame (the set of canonical variables) reached by the Hamiltonian flow of the cubic generator, and by that
flow alone, the Hamiltonian contains no term in which a single wavenumber
carries the minority sign, at any order. The vanishing of the non-trivial
four-wave interactions is the quartic instance, the fifth-order zeros of the
literature are the quintic one, and the reason is a property specific to planar
deep water: a harmonic field splits exactly into two holomorphic functions,
of $x-\mathrm{i}y$ and $x+\mathrm{i}y$, carrying the positive and the
negative wavenumbers respectively. The quadratic pressure term is the product of the two, so pressure alone
cannot drive a process in which a single wavenumber carries the minority
sign; such processes enter only through the motion of the surface on which
the potential is evaluated, and the exact flow of the cubic generator removes
those kinematic effects completely.

\section{Real canonical variables and the truncated Hamiltonian}\label{sec:setup}
 
We work in the dimensionless deep-water setting of \citet{Zakharov1968} with one horizontal dimension,
in which the surface elevation $\eta(x,t)$ and the surface potential
$\phi(x,t)$ are canonically conjugate, $\eta_t=\delta\mathcal{H}/\delta\phi$,
$\phi_t=-\delta\mathcal{H}/\delta\eta$. The Hamiltonian admits the standard expansion in the
steepness-like parameter $\epsilon$ as
\begin{multline}\label{eq:H_etaphi}
\mathcal{H}^{\mathrm{phys}}[\eta,\phi]=\frac12\!\int(\eta^2+\phi H\phi_x)\,\mathrm{d}x
+\frac{\epsilon}{2}\int \left[(\phi_x)^2-(H\phi_x)^2\right]\eta\,\mathrm{d}x\\
+\frac{\epsilon^2}{2}\!\int\big[\eta^2\phi_{xx}H\phi_x
+\phi H(\eta H(\eta H\phi_x)_x)_x\big]\mathrm{d}x+O(\epsilon^3),
\end{multline}
where $H$ is the Hilbert transform with Fourier symbol $-\ii\sgn(k)$,
subscripts denote partial derivatives, and integrals are over the real line. Truncating the Hamiltonian at fourth order in the fields, two orders in $\epsilon$ beyond linear theory, is the standard starting
point of the four-wave theory of deep-water gravity waves, most notably for deriving the
reduced Zakharov equation \citep{Zakharov1968,Krasitskii1994}.

We adopt the real canonical variables $(u,v)$
of \citet{Simson2026}, defined by the near-identity transformation
\begin{align}\label{eq:can_eta}
    \eta &= u - \epsilon\, H(Hu\, Hu_x)
         - \tfrac{1}{2}\,\epsilon^2 H((Hu)^2 Hu_x)_x + O(\epsilon^3),\\
\label{eq:can_phi}
    \phi &= v - \epsilon\, H(Hu\, Hv_x)
         - \tfrac{1}{2}\,\epsilon^2 H((Hu)^2 Hv_x)_x + O(\epsilon^3).
\end{align}
The transformation is canonical through $O(\epsilon^2)$ and carries no independent quartic generator: through this order it is the Taylor expansion of the Hamiltonian flow of the cubic generator, the unique functional
    $F[u,v]=-\tfrac{1}{2}\int(Hu)^2Hv_x\,\mathrm{d}x$
that eliminates the cubic Hamiltonian \citep[see for example][and references therein]{CraigGuyenneSulem2021}, so everything it affects at quartic order (and beyond) descends from the cubic terms alone. A detailed account of the exact flow is given in \S\ref{exact_flow_section}.

Substituting
\eqref{eq:can_eta}--\eqref{eq:can_phi} into \eqref{eq:H_etaphi} removes the
cubic terms and, after
simplification, gives
\begin{multline}\label{eq:ham_uv}
\mathcal{H}[u,v]=\frac{1}{2}\int (u^2+vHv_x)\,\mathrm{d}x\\
-\epsilon^2\int\Big(v_{xx}H(uHu)+\tfrac{1}{2}HuH(HuHv_x)_x-\tfrac{1}{2}uH(uHv_x)_x\Big)Hv_x\,\mathrm{d}x
+O(\epsilon^3)\\
\equiv \mathcal{H}_2+\epsilon^2\mathcal{H}_4+O(\epsilon^3).
\end{multline}
Notably, the effect of the transformation on the quartic Hamiltonian can be summarized as 
\begin{equation}\label{effect}
\mathcal{H}_4[u,v]=\mathcal{H}^{\mathrm{phys}}_4[u,v]
-\mathcal{H}^{\mathrm{phys}}_4[Hu,v]
\equiv\mathrm{Re}\,\mathcal{H}^{\mathrm{phys}}_4[u+\ii Hu,v],
\end{equation}
where $\mathcal{H}_4^{\mathrm{phys}}[\eta,\phi]$ is the quartic part of
\eqref{eq:H_etaphi}. Essentially the same identity was obtained by
\citet{CraigSulem2016} as part of an analysis of the mapping properties of the elimination of
the cubic and the non-resonant quartic terms in Sobolev spaces. It was later used by
\citet{CraigGuyenneSulem2021} to derive their Hamiltonian Dysthe equation,
the subtracted term supplying the mean-flow contribution. This identity,
resulting from the cubic elimination alone, is particular to deep water.

The equations of motion following from \eqref{eq:ham_uv}, $u_t=\delta\mathcal{H}/\delta v$,
$v_t=-\delta\mathcal{H}/\delta u$, are given explicitly in \citet{Simson2026}.
A monochromatic wave is an exact solution of these equations, and the transformation turns it into the third-order Stokes wave. The reason is given in \S\ref{sec:order_by_order}.

The conservation laws for mass, momentum, and energy in $(u,v)$ variables are, respectively,
\begin{equation}\label{conservation_laws}
    \int u\, \mathrm{d}x=\mathrm{const}, \quad \int uv_x\, \mathrm{d}x =\mathrm{const}, \quad \mathcal{H}=\mathrm{const}.
\end{equation}
These correspond to three symmetries of the Hamiltonian, i.e., transformations under which the Hamiltonian is invariant: the shift of the potential $v\mapsto v+v_0$, the spatial translation $x\mapsto x+x_0$, and the translation of time $t\mapsto t+t_0$, respectively.

\section{Additional symmetry and its conservation law}\label{sec:symmetry_and_W}

The Hamiltonian \eqref{eq:ham_uv} is also
invariant under the transformation
\begin{equation}\label{eq:hilbrot}
    (u,v)\mapsto(\mathrm{e}^{\alpha H}u,\;\mathrm{e}^{\alpha H}v),\qquad \alpha\in\mathbb{R},
\end{equation}
which we call the Hilbert rotation. Since $H^2=-1$, the exponential form simplifies to $\mathrm{e}^{\alpha H}=\cos(\alpha)+\sin(\alpha)H$.
The symmetry becomes transparent upon introducing the analytic signals 
\begin{equation}
    p=u+\ii Hu,\qquad q=|\partial_x|(v+\ii Hv),
\end{equation}
with $|\partial_x|\equiv H\partial_x$, which transform simply as 
\begin{equation}
    (p,q)\mapsto\mathrm{e}^{-\ii\alpha}\,(p,q).
\end{equation}
By \eqref{effect}, $\mathcal{H}_4=\mathrm{Re}\,\mathcal{H}^{\mathrm{phys}}_4[p,v]$. Writing
$|\partial_x|v=\tfrac12(q+\bar q)$ in it, the terms with $qq$ contain positive wavenumbers
only and integrate to zero, those with $q\bar q$ cancel after integration by parts, and
the Hamiltonian \eqref{eq:ham_uv} becomes
\begin{equation}\label{eq:H_pq}
  \mathcal{H}[p,q]=\frac14\!\int\!\big(p\bar p+\bar q\,|\partial_x|^{-1}q\big)\,\mathrm{d}x
  -\frac{\epsilon^2}{16}\,\mathrm{Re}\!\int\!\Big[\,p^{2}\,|\partial_x|\,\bar q^{\,2}
  -2\,(p\bar q)\,|\partial_x|\,(p\bar q)\Big]\,\mathrm{d}x ,
\end{equation}
where the overlines denote complex conjugation, and the invariance is manifest.

At first order in $\alpha$, the transformation is $(u,v)\mapsto(u+\alpha Hu, v+\alpha Hv)$, so the maps \eqref{eq:hilbrot} are the flow of the vector field $(Hu,Hv)$ on phase space.
The flow equations are $\partial_\alpha u=Hu$, $\partial_\alpha v=Hv$. These can be put into Hamiltonian form $\partial_\alpha u=\delta W/\delta v$, $\partial_\alpha v=-\delta W/\delta u$ with the functional
\begin{equation}\label{eq:W}
    W[u,v]=\int vHu\,\mathrm{d}x,
\end{equation}
since $\delta W/\delta v=Hu$ and $-\delta W/\delta u=Hv$ by the skewness of $H$. The
Hilbert rotation is therefore the Hamiltonian flow of $W$.

The invariance of $\mathcal{H}$ along this flow is now equivalent to the
conservation of $W$ by the dynamics (Noether's theorem):
\begin{equation}\label{eq:Wcons}
    \frac{\mathrm{d}\mathcal{H}}{\mathrm{d}\alpha}=\{\mathcal{H},W\}=-\{W,\mathcal{H}\}=
    -\frac{\mathrm{d}W}{\mathrm{d}t}=0,
\end{equation}
where $\{A,B\}=\int(\delta A/\delta u\,\delta B/\delta v
-\delta A/\delta v\,\delta B/\delta u)\,\mathrm{d}x$ 
is the canonical bracket. In this sense $W$ is the Noether charge of the Hilbert rotation.

The physical interpretation of the conservation law is clearest at linear order,
where the field splits exactly into left- and right-moving waves,  
\begin{equation}\label{eq:split}
    u=l+r,\qquad v=A^{-1}(r-l),\qquad l=\tfrac12(u-Av),\qquad r=\tfrac12(u+Av),
\end{equation} 
where $A=-H|\partial_x|^\frac{1}{2}$, with symbol $\ii \sgn(k)\sqrt{|k|}$. In these variables, the quadratic Hamiltonian splits as
$\mathcal{H}_2[l,r]=\int l^2\,\mathrm{d}x+\int r^2\,\mathrm{d}x$, and the fields evolve independently:  $l_t=Al$ and $r_t=-Ar$. $W$ becomes
\begin{equation}\label{eq:Wsplit}
    W[l,r]=\int r|\partial_x|^{-\frac{1}{2}}r\,\mathrm{d}x
     -\int l|\partial_x|^{-\frac{1}{2}}l\,\mathrm{d}x=N^{+}-N^{-}.
\end{equation}
The functionals $N^{+}$ and $N^{-}$ are the wave actions (or numbers of waves) of the
right- and left-moving components. In Fourier space,
$N^{+}=\int|\hat r(k)|^{2}\,|k|^{-1/2}\,\mathrm{d}k$, where the integrand is the spectral energy density
$|\hat r(k)|^{2}$ divided by the frequency $\sqrt{|k|}$, and likewise for $N^{-}$. The conservation of $W$ is therefore the conservation of the wave-action imbalance between the right- and left-moving waves:
\begin{equation}\label{eq:imbalance}   W=\int vHu\, \mathrm{d}x=N^+ -N^-=\mathrm{const}.
\end{equation}

In physical terms, \eqref{eq:imbalance} means that right-moving waves cannot be converted into left-moving waves, nor vice versa. The two populations can change only together and by equal amounts, $\Delta N^+=\Delta N^-$.

Beyond linear order the split \eqref{eq:split} no longer separates the two
directions of propagation exactly. The quartic Hamiltonian
$\mathcal{H}_4[u,v]$ does not conserve the total wave action
$N=N^{+}+N^{-}$, and since $W$ remains constant, neither $N^{+}$ nor
$N^{-}$ is conserved on its own. The identity \eqref{eq:Wsplit} is
algebraic and unaffected, so \eqref{eq:imbalance} remains exact, while
$N^{+}$ and $N^{-}$, as defined in \eqref{eq:Wsplit}, represent the
unidirectional wave actions only approximately. The nonlinearity
thus
alters the actions individually, by amounts of order $\epsilon^2$, but leaves their difference untouched.

In physical variables the charge acquires higher-order corrections.
Inverting \eqref{eq:can_eta}--\eqref{eq:can_phi} and inserting into
\eqref{eq:W} gives
\begin{multline}\label{eq:W_phys}
W=\int\phi H\eta\,\mathrm{d}x+\frac{\epsilon}{2}\int\eta^{2}\phi_x\,\mathrm{d}x\\
+\frac{\epsilon^{2}}{2}\int\Big[\eta\phi_x H(\eta^{2})_x
-\eta^{2}\phi_x H\eta_x+\eta^{2}\eta_x H\phi_x\Big]\mathrm{d}x
+O(\epsilon^{3}),
\end{multline}
which coincides through $O(\epsilon^{2})$ with the first nonlocal charge
of \citet{Neveu2021}, obtained there by direct perturbative calculation.

\section{Four-wave kernel and wavenumber-sign sectors}\label{sec:kernel_and_signs}

\subsection{Four-wave kernel}\label{sec:kernel}

We recall the four-wave kernel and its reduction on the momentum shell $\sum_i k_i=0$ from \citet{Simson2026}, with added details, since everything below depends on them.

In Fourier space, with the symmetric convention
$\hat{f}(k)=(2\pi)^{-1/2}\int f(x)\,\mathrm{e}^{-\ii k x}\,\mathrm{d}x$, the
quartic part of the Hamiltonian \eqref{eq:ham_uv} is written as
\begin{equation}\label{eq:H4_fourier}
  \mathcal{H}_4 = -\frac{1}{2\pi} \!\int\! K_{k_1k_2k_3k_4}\,
  \hat{u}(k_1)\hat{u}(k_2)\hat{v}(k_3)\hat{v}(k_4)\,\delta\!\Big(\sum_i k_i\Big)
  \prod_i \mathrm{d}k_i,
\end{equation}
where $K_{k_1k_2k_3k_4}\equiv K(k_1,k_2,k_3,k_4)$ is the quartic kernel describing four-wave interactions. Its first two arguments are the $u$ slots and the last two the $v$ slots (the $\eta$ and $\phi$ slots for a kernel in physical variables).
A direct evaluation, symmetrized in $k_1\leftrightarrow k_2$ and
$k_3\leftrightarrow k_4$, yields 
\begin{equation}\label{eq:K-closed}
  K_{k_1k_2k_3k_4} = \tfrac{1}{8}|k_3k_4|\,(1+\sigma_1\sigma_2)
  \!\left[\,2(|k_3|+|k_4|) \;-\sum_{i=1}^{2}\sum_{j=3}^{4}|k_i+k_j|\,\right],
\end{equation}
where $\sigma_i \equiv \sgn(k_i)$. Note that this form has no frequency denominators that come with the elimination of cubic terms in the complex normal variable formulation. Here none arises at any stage, and the result is a polynomial in
$|k_i|$, $\sgn (k_i)$ and $|k_i+k_j|$.

When compared to the original quartic kernel of $\mathcal{H}^{\mathrm{phys}}_4$ in \eqref{eq:H_etaphi}, we get the relation
\begin{equation}\label{K_K_phys}
    K=(1+\sigma_1\sigma_2)K^{\mathrm{phys}}.
\end{equation}
In $x$-space this is the identity \eqref{effect}. The factor $(1+\sigma_1\sigma_2)$ makes the
kernel $K$ vanish unless $\sigma_1=\sigma_2$.

On the momentum shell we have $|k_1+k_4|=|k_2+k_3|$ and $|k_2+k_4|=|k_1+k_3|$, so \eqref{eq:K-closed} reduces to 
\begin{equation}\label{eq:K-shell}
  K_{k_1k_2k_3k_4}=\tfrac14(1+\sigma_1\sigma_2)\,|k_3k_4|\,
  \Big[\,|k_3|+|k_4|-|k_1+k_3|-|k_2+k_3|\,\Big],
\end{equation}
symmetric under $k_3\leftrightarrow k_4$ by the same identities. With $\sigma_1=\sigma_2$  the square bracket vanishes if $\sigma_3=-\sigma_4$, whereas for $\sigma_3=\sigma_4$ it equals $2\min_i\vert{}k_i\vert{}$ by the min identity of \citet{DyachenkoKachulinZakharov2017}. Hence on the momentum shell the kernel \eqref{eq:K-closed}
evaluates to
\begin{equation}\label{eq:K-momentum}
  K_{k_1k_2k_3k_4} \;=\;
  \begin{cases}
    |k_3k_4|\,\min(|k_1|,|k_2|,|k_3|,|k_4|), & \sigma_1=\sigma_2=-\sigma_3=-\sigma_4, \\[2pt]
    0,& \text{otherwise.}
  \end{cases}
\end{equation}

The fact that $K^{\mathrm{phys}}$ vanishes
when a single wavenumber of the opposite sign occupies a $\phi$ slot, $\sigma_1=\sigma_2$,
$\sigma_3=-\sigma_4$, was noted by \citet{Ussembayev2019}, who shows it for
the physical kernels of every order, not just for the quartic one. For $\sigma_1=-\sigma_2$, by contrast,
$K^{\mathrm{phys}}$ is in general non-zero, and it is the factor
$(1+\sigma_1\sigma_2)$ in \eqref{K_K_phys}, produced by the cubic elimination, that removes that part from $K$.
Together the two facts make $K$ vanish whenever a single wavenumber has the opposite sign
from the other three. This vanishing contains DKZ's sign rule
(\S\ref{sec:zakharov_section}) and is the quartic instance of the selection rule established
in \S\ref{sec:origin}.

With the momentum shell form \eqref{eq:K-momentum}, the invariance of $\mathcal{H}_4$ under the Hilbert rotation \eqref{eq:hilbrot} becomes obvious. In Fourier space the transformation acts as \begin{equation}
    \hat{u}(k_i)\mapsto\mathrm{e}^{-\ii\alpha\sigma_i}\hat{u}(k_i),\qquad \hat{v}(k_i)\mapsto\mathrm{e}^{-\ii\alpha\sigma_i}\hat{v}(k_i)
\end{equation} in \eqref{eq:H4_fourier}, and the integrand acquires a total factor $\mathrm{e}^{-\ii\alpha\sum_i\sigma_i}=1$, because the sign sum $\sum_i\sigma_i=0$ on the kernel's support. Hence $\mathcal{H}_4$ does not change under the Hilbert rotation. The same holds for the quadratic Hamiltonian $\mathcal{H}_2=\frac{1}{2}\int (|\hat{u}(k)|^2+|k||\hat{v}(k)|^2)\,\mathrm{d}k$, and so for $\mathcal{H}=\mathcal{H}_2+\epsilon^2\mathcal{H}_4$ also.

\subsection{Wavenumber-sign sectors}\label{sec:sectors}

For the kernel of the $n$-th order Hamiltonian $\mathcal{H}_n$, we say that a point
$(k_1,\dots,k_n)$ in the $n$-dimensional wavenumber space, satisfying the momentum
condition $\sum_i k_i=0$, belongs to the \emph{$(n-m)$-$m$ wavenumber-sign
sector}, $0\le m\le n/2$, if $m$ of the signs $\sigma_i=\operatorname{sgn}k_i$
take one value and $n-m$ the other. Sectors are unoriented, the two sign
values play symmetric roles, and the larger count is written first. 

These sectors refer strictly to wavenumber signs and must not be confused with the interaction classes of the complex normal variable description, such as the $2\leftrightarrow2$, $3\leftrightarrow1$, and $4\leftrightarrow0$ four-wave processes (here, the last two change the number of waves). In complex normal variables the kernel $K$
generates every four-wave interaction class at once, and its $2\leftrightarrow2$ ($\bar a\bar a aa$) class is the reduced Zakharov kernel in the closed form of \citet{Dyachenko2020canonical}, as shown in \S\ref{sec:zakharov_section}.

In general, under the Hilbert rotation the integrand of any term of
the Hamiltonian acquires the factor $\mathrm{e}^{-\ii\varsigma\alpha}$, where the exponent
\begin{equation}
    \varsigma=\sum_i\sigma_i
\end{equation} is the sum of the slot signs. The sum is constant on each wavenumber-sign sector,
and we say that the sector carries charge $\varsigma$. For a part $\mathcal{H}^{(\varsigma)}$ of the Hamiltonian with charge $\varsigma$, the derivative with respect to $\alpha$ is $\{\mathcal{H}^{(\varsigma)},W\}$, as in
\eqref{eq:Wcons}, and since $\partial_\alpha(\mathrm{e}^{-\ii\varsigma\alpha})=-\ii\varsigma \mathrm{e}^{-\ii\varsigma\alpha}$, we have the equation
\begin{equation}\label{eq:charge}
    \{\mathcal{H}^{(\varsigma)},W\}=-\ii\varsigma \mathcal{H}^{(\varsigma)}.
\end{equation}
The Hilbert rotation is therefore a symmetry exactly when the parts with $\varsigma\neq0$ vanish.
In general, a part with $\varsigma\neq0$ contributes $\ii\varsigma\mathcal{H}^{(\varsigma)}$ to
$\mathrm{d}W/\mathrm{d}t$.

For the quartic kernel $K(k_1,k_2,k_3,k_4)$ only the 2-2 and 3-1 sectors occur,
since momentum conservation $\sum_i k_i=0$ excludes the $n$-0 sector for every $n$. The charge is $0$ on 2-2 (neutral sector) and $\pm2$ on 3-1 (charged sector).
The \emph{aligned} part of the 2-2 sector is the subset with
$\sigma_1=\sigma_2=-\sigma_3=-\sigma_4$, the two $u$-wavenumbers sharing one sign
and the two $v$-wavenumbers the other; the remaining two pairings,
$\sigma_1=\sigma_3=-\sigma_2=-\sigma_4$ and $\sigma_1=\sigma_4=-\sigma_2=-\sigma_3$,
each pair a $u$- with a $v$-wavenumber and form the \emph{non-aligned} part.

Note that $\mathcal{H}_4$ is invariant
under the Hilbert rotation if and only if its symmetrized kernel is zero outside the
\mbox{2-2} wavenumber-sign sector. The Hilbert rotation symmetry is therefore a statement about where the kernel vanishes. The support \eqref{eq:K-momentum} is narrower than the symmetry requires, being confined to the aligned part of the \mbox{2-2} sector, but for a Hamiltonian of the form \eqref{eq:ham_uv} the two facts
are not independent. Writing the three quartic terms with coefficients
$c_1$, $\tfrac12 c_2$, $-\tfrac12 c_3$ in place of
$1$, $\tfrac12$, $-\tfrac12$, it can be verified that the kernel vanishes on the 3-1 sector if
and only if $c_1=c_2=c_3$, and under the same condition it vanishes on the non-aligned 2-2 sector as well. The cubic elimination in deep water produces the three quartic terms in exactly the proportion
that makes their 3-1 and non-aligned 2-2 parts cancel. Equation \eqref{effect} expresses the same fact, and shows the mechanism in $x$-space.

\section{Vanishing of non-trivial resonant interactions}\label{sec:vanishing_demonstration}

In real variables, each mode oscillates at both frequencies $\pm\omega_i$, where \mbox{$\omega_i\equiv\sqrt{\vert{}k_i\vert{}}$}. A quartic term is resonant when $\sum_{i=1}^4 s_i\,\omega_i=0$ for some sign assignment $s_i=\pm 1$. On the momentum shell, this sum cannot vanish for sign assignments with
four or three equal signs, leaving only two-against-two sign configurations. This means that the resonant interactions are only of the $2\leftrightarrow2$ class: two waves scattering into two. The resonant manifold $\mathcal{R}$ is therefore the three-sheeted surface defined by
\begin{equation}\label{eq:resonance_conditions}
    k_1+k_2+k_3+k_4=0,\qquad \gamma\equiv\Delta_1\Delta_2\Delta_3=0,
\end{equation}
where
\begin{equation}
    \begin{aligned}
\Delta_1&=\omega_1+\omega_2-\omega_3-\omega_4,\\
\Delta_2&=\omega_1-\omega_2+\omega_3-\omega_4,\\
\Delta_3&=\omega_1-\omega_2-\omega_3+\omega_4.
\end{aligned}
\end{equation}

The trivial solutions of \eqref{eq:resonance_conditions} are the
configurations in which the wavenumbers cancel in pairs, one condition
for each of the three pairings,
\begin{equation}\label{eq:trivial}
    k_1+k_2=k_3+k_4=0,\qquad k_1+k_3=k_2+k_4=0,\qquad k_1+k_4=k_2+k_3=0 ,
\end{equation}
any one of which defines a plane of the momentum shell on which two of
the $\Delta_i$ vanish identically. These planes form the trivial part of $\mathcal R$, on which the resonant four-wave interaction only renormalizes the frequencies (Stokes-type corrections) and in whose neighbourhood it produces the modulational dynamics.

The remaining non-trivial solutions form the non-trivial part of $\mathcal{R}$: 
\begin{equation}\label{eq:NTzeta}
\begin{aligned}
    k_1&=a\,(1+\zeta)^2,\\
    k_2&=a\,\zeta^2(1+\zeta)^2,\\
    k_3&=a\,\zeta^2,\\
    k_4&=-a\,(1+\zeta+\zeta^2)^2,
\end{aligned}
\qquad a\in\mathbb{R}\setminus\{0\},\quad 0<\zeta\leqslant1,
\end{equation}
together with all permutations of the four entries. This is the parametrization of \citet{DyachenkoZakharov1994} transcribed
to the slots of the real kernel (their $k_2$ and $k_3$ entering here as
$-k_3$ and $-k_4$).
In the labelling of \eqref{eq:NTzeta} the only resonant combination is
$\omega_1+\omega_2=\omega_3+\omega_4$, with the momentum condition in the same
arrangement $k_1+k_2=(-k_3)+(-k_4)$. Physically, for $a>0$, \eqref{eq:NTzeta} represents the resonant $(k_1,k_2)\rightarrow(-k_3,-k_4)$ interaction between a right-moving pair of waves of wavenumbers $\vert{}k_1\vert{}$ and $\vert{}k_2\vert{}$, and a second pair formed by a left-moving wave of wavenumber $-\vert{}k_3\vert{}$ (the longest of the four) and a right-moving wave of wavenumber $\vert{}k_4\vert{}$ (the shortest).

Although kinematically allowed, these interactions are dynamically forbidden: every non-trivial resonance \eqref{eq:NTzeta} has three
wavenumbers of one sign and one of the other, and therefore lies in the
3-1 wavenumber-sign sector, where the real kernel \eqref{eq:K-momentum}
vanishes identically. In complex normal variables the same fact appears as the cancellation of \citet{DyachenkoZakharov1994}, where dozens of terms of the reduced Zakharov kernel sum to zero on the non-trivial resonant manifold. In the present variables there is nothing left to cancel: with \eqref{eq:K-momentum} the same zero follows by inspection. The on-manifold value of the four-wave coefficient is invariant under near-identity canonical transformations, so the conclusion itself does not depend on the variables.

The exact set of dynamically active resonances, those on which the kernel \eqref{eq:K-momentum} does not vanish, can be obtained by adding
the support condition,
\begin{equation}
    \sigma_1=\sigma_2=-\sigma_3=-\sigma_4,\qquad \sigma_i\equiv\sgn(k_i),
\end{equation}
to the system \eqref{eq:resonance_conditions}. The combined system is easy to solve. The momentum condition becomes $|k_1|+|k_2|=|k_3|+|k_4|$ and the resonance condition \mbox{$\gamma=0$} collapses to $|k_1k_2|=|k_3k_4|$, giving only the trivial quartets 
\begin{equation}
    \{|k_1|,|k_2|\}=\{|k_3|,|k_4|\}
\end{equation}
 as the solution. This is a subset of the trivial resonant manifold \eqref{eq:trivial}: only the planes
$k_1+k_3=0$ and $k_1+k_4=0$ within the support, the third plane
$k_1+k_2=0$ lying outside it with the two $u$-wavenumbers of opposite
sign.

\begin{figure}
  \centerline{\includegraphics[width=\textwidth]{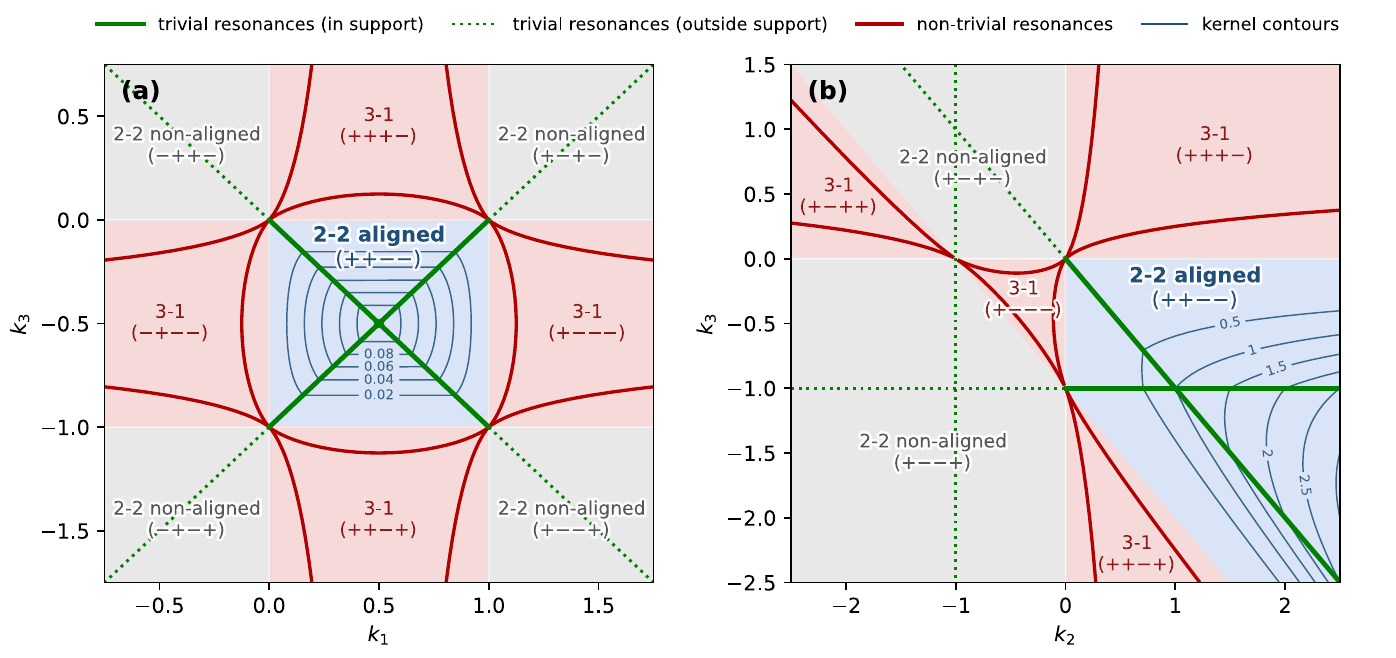}}
  \caption{Wavenumber-sign sectors of the four-wave kernel \eqref{eq:K-momentum} on two slices of the
momentum shell, each region labelled by its sign pattern
$(\sigma_1\sigma_2\sigma_3\sigma_4)$. Blue: the aligned 2-2 sector, the
support of $K$, with contours of $K$ labelled by their value on the slice;
red: the 3-1 sector, $K=0$; grey: the non-aligned 2-2 sector, $K=0$. Green lines are trivial resonances, solid inside the support and dotted outside it; red curves are the non-trivial resonances, which lie entirely in the 3-1 sector and meet the kernel's support only at its corners, where two wavenumbers vanish. 
(a) The slice $k_1+k_2=1$,
(b) The slice $k_1=1$.
}
  \label{fig:sectors}
\end{figure}

The geometry can be visualized. Every condition defining $\mathcal{R}$ and
the sectors is homogeneous in $(k_1,k_2,k_3,k_4)$, so a non-central slice of the momentum shell $k_1+k_2+k_3+k_4=0$  shows the whole
structure up to scaling. Figure~\ref{fig:sectors} shows two such slices, with green lines and red curves representing the trivial and non-trivial parts of the resonant manifold $\mathcal{R}$, respectively. The kernel \eqref{eq:K-momentum} is non-zero only in the blue region. 
In figure~\ref{fig:sectors}(a) we have $k_1+k_2=1$ (hence $k_3+k_4=-1$), with $k_1$ and $k_3$ as coordinates, and the support of the kernel is the blue square $0<k_1<1$, $-1<k_3<0$; its mirror image $k_1+k_2=-1$ is identical by the reality of the fields, and the third trivial plane, $k_1+k_2=0$, does not meet the slice. In figure~\ref{fig:sectors}(b) we instead have $k_1=1$ (hence $k_2+k_3+k_4=-1$), with $k_2$ and $k_3$ as coordinates. The support is now an unbounded wedge and all resonance curves are open; the third trivial plane is visible as the vertical dotted line. On both slices the non-trivial curves lie entirely in the 3-1 sector, where the kernel vanishes, and the trivial lines pass through the support, where the kernel is non-zero. 
The vanishing of \citet{DyachenkoZakharov1994} is a statement about the
red curves: on those curves the kernel vanishes. DKZ's sign rule
is a statement
about the red strips:
both the real-variable kernel and the reduced Zakharov kernel are zero there.

\section{Relation to the Zakharov equation}\label{sec:zakharov_section}
We now introduce the complex normal variable $a_k$ through the usual relations
\begin{equation}\label{normal_variables}
    \hat u(k)=\sqrt{\frac{\omega}{2}}\,(a_k+\bar a_{-k}),\qquad
    \hat v(k)=-\frac{\ii}{\sqrt{2\omega}}\,(a_k-\bar a_{-k}),\qquad
    \omega=\sqrt{|k|}.
\end{equation}
Direct substitution into \eqref{eq:H4_fourier}, with $k\to-k$ for every
$\bar a$ factor and symmetrization within each interaction class ($aaaa$, $\bar aaaa$ etc.), gives
the coefficients of all classes. Written as
$\tfrac12\int T_{kk_1k_2k_3}\,\bar a_k\bar a_{k_1}a_{k_2}a_{k_3}\,
\delta(k+k_1-k_2-k_3)\,\mathrm{d}k\,\mathrm{d}k_1\mathrm{d}k_2\mathrm{d}k_3$, the $\bar a\bar a aa$ ($2\leftrightarrow2$) class  of $\mathcal{H}_4$ has the coefficient 
\begin{align}\label{T_aaaa_full}
T_{kk_1k_2k_3}
 &= \frac{1}{4\pi\,|kk_1k_2k_3|^{\frac14}}
    \Big[\,|kk_1|^{\frac12}\,K_{-k,-k_1,k_2,k_3}
        + |k_2k_3|^{\frac12}\,K_{k_2,k_3,-k,-k_1} \nonumber\\
 &\qquad\qquad\qquad
        - |kk_2|^{\frac12}\,K_{-k,k_2,-k_1,k_3}
        - |kk_3|^{\frac12}\,K_{-k,k_3,-k_1,k_2} \nonumber\\
 &\qquad\qquad\qquad
        - |k_1k_2|^{\frac12}\,K_{-k_1,k_2,-k,k_3}
        - |k_1k_3|^{\frac12}\,K_{-k_1,k_3,-k,k_2}\Big].
\end{align}
This is the only class the reduced Zakharov equation keeps. On co-propagating quartets (all four wavenumbers of one sign) only the first two terms lie on the support of $K$ \eqref{eq:K-momentum}, and \eqref{T_aaaa_full} reduces to
\begin{equation}\label{eq:T_co}
  T_{kk_1k_2k_3}=\frac{1}{4\pi}\,|kk_1k_2k_3|^{1/4}
  \Big(\sqrt{|kk_1|}+\sqrt{|k_2k_3|}\Big)\min(|k|,|k_1|,|k_2|,|k_3|),
\end{equation}
which is the unidirectional Zakharov kernel of \citet[eq.~3.5]{DyachenkoKachulinZakharov2017}.
On quartets with one negative wavenumber in each pair, say
$k,k_2>0$ and \mbox{$k_1,k_3<0$}, only the terms with $u$-slot pairs $(-k, k_3)$ and $(-k_1, k_2)$ survive: the overall sign of \eqref{eq:T_co} is reversed and the second factor becomes
$\big(\sqrt{|kk_3|}+\sqrt{|k_1k_2|}\big)$. On the remaining quartets, for which $kk_1k_2k_3<0$, every
contribution to the $\bar a\bar a aa$ class comes from outside the
support of $K$ and gives \mbox{$T\equiv0$}. Together the three cases are exactly the closed form of the reduced Zakharov kernel of
\citet[eq.~6]{Dyachenko2020canonical}, including DKZ's sign rule ($T\equiv0$ for $kk_1k_2k_3\leq0$):
\begin{equation}\label{zakharov_kernel_d2020}
    T_{kk_1k_2k_3}=\frac{1}{4\pi}\,|kk_1k_2k_3|^{\frac14}\,m\times
\begin{cases}
\;\;\;|kk_1|^{\frac12}+|k_2k_3|^{\frac12}, & \sigma_k=\sigma_1=\sigma_2=\sigma_3,\\[2pt]
-\big(|kk_3|^{\frac12}+|k_1k_2|^{\frac12}\big), & \sigma_k=\sigma_2=-\sigma_1=-\sigma_3,\\[2pt]
-\big(|kk_2|^{\frac12}+|k_1k_3|^{\frac12}\big), & \sigma_k=\sigma_3=-\sigma_1=-\sigma_2,\\[2pt]
\;\;\;0, & kk_1k_2k_3\leq0,
\end{cases}
\end{equation}
where $m=\min{(|k|, |k_1|, |k_2|, |k_3|)}$, $\sigma_k=\sgn(k)$ and $\sigma_j=\sgn(k_j)$.

The fact that \eqref{T_aaaa_full} coincides with the
reduced Zakharov kernel, although the variables \eqref{normal_variables} differ from
those of that equation by the quartic canonical transformation, can be explained. The cubic generator that
removes $\mathcal{H}_3$ is unique, deep water having no three-wave
resonances, and the quartic generator that produces the reduced Zakharov
Hamiltonian by removing the other classes, taken without an
$\bar a\bar a aa$ part of its own, changes only those classes at quartic order and leaves
$\bar a\bar a aa$ untouched. Such a part is not fixed by the reduction: it changes the
$\bar a\bar a aa$ coefficient by a multiple of
$\omega(k)+\omega(k_1)-\omega(k_2)-\omega(k_3)$, so the reduced kernel is defined off the
resonant manifold only up to this choice, and \eqref{zakharov_kernel_d2020} is the one
obtained without it. The reduced Zakharov kernel is thus a
projection of the elementary kernel \eqref{eq:K-closed}, and its sign
rule is the symmetry of \S\ref{sec:symmetry_and_W} seen in one
interaction class. With \eqref{T_aaaa_full}, the main argument in the Zakharov frame becomes reversed: the vanishing on the non-trivial resonant manifold now simply follows from it, rather than being the foundation for obtaining the closed form \eqref{zakharov_kernel_d2020}.

With \eqref{normal_variables} the charge becomes 
\begin{equation}\label{normal_W}
W=\int\sgn(k)\,|a_k|^{2}\,\mathrm{d}k
 =\int_0^\infty|a_k|^{2}\,\mathrm{d}k-\int_{-\infty}^0|a_k|^{2}\,\mathrm{d}k
 =N^{+}-N^{-}.
\end{equation}
In normal variables the Hilbert rotation \eqref{eq:hilbrot} is
$a_k\mapsto\mathrm{e}^{-\ii\alpha\,\sgn(k)}a_k$, while the total action $N$ generates the uniform
rotation $a_k\mapsto\mathrm{e}^{-\ii\alpha}a_k$. A monomial
$\bar a_k\bar a_{k_1}a_{k_2}a_{k_3}$ with $k+k_1=k_2+k_3$ is invariant under the first unless
$kk_1k_2k_3<0$, so for this class the invariance is DKZ's sign rule.

Equation \eqref{normal_W} has the form of the additional invariants of
\citet{ZAKHAROV1988283}, $I=\int f_k\,|a_k|^2\,\mathrm{d}k+\dots$, with
$f_k=\sgn (k)$ and no correction terms. Such an invariant exists only if on each resonance either $f$ is
conserved, $f_k+f_{k_1}=f_{k_2}+f_{k_3}$, or the four-wave coefficient
vanishes (their theorem~3.1). The sign is conserved on the trivial part
of the resonant manifold, where the wavenumbers cancel in pairs, and not on
the non-trivial part, where three of them share a sign. The theorem thus
ties the existence of $W$ to the vanishing there, and both are
consequences of the same symmetry. 

The reduced Zakharov equation, in its own variable $b_k$, conserves the
total action $N=\int|b_k|^2\,\mathrm{d}k$ by the uniform phase invariance, and $W=\int\sgn(k)|b_k|^2\,\mathrm{d}k$
by the Hilbert rotation. Therefore it also conserves $N^{+}$ and $N^{-}$ separately, as in
\citet{Dyachenko2019fluids,Dyachenko2020canonical}. The present frame
retains the number-changing classes, so here $W$ is conserved while $N$
is not.

\section{Origin of the symmetry}\label{sec:origin}

So far we have studied only the quartic truncation of the deep-water Hamiltonian. We observed that once the cubic terms are eliminated, it is invariant under the Hilbert rotation, and the invariance accounts for the vanishing of the non-trivial resonant interactions. At the kernel level, it is the vanishing of the whole 3-1 wavenumber-sign sector. This is a description of the structure, but it does not explain it, and it does not say whether anything of it survives beyond the quartic order.
In this section we take a wider view, the Hamiltonian at every order and the transformation \eqref{eq:can_eta}--\eqref{eq:can_phi} completed to an exact flow, and we trace the symmetry to more general properties of deep-water theory.

\subsection{Decomposition into one-sided fields}\label{sec:decomposition}

We split the physical fields into their positive- and negative-wavenumber parts,
\begin{equation}
    \eta=\eta_+ + \eta_-,\qquad \phi=\phi_+ + \phi_-,
\end{equation}
and regard the four components as independent variables. On real fields $\eta_-=\bar\eta_+$ and
$\phi_-=\bar\phi_+$. In Fourier space, the canonical bracket pairs opposite wavenumbers:
$\{\hat\eta(k),\hat\phi(k')\}=\delta(k+k')$ is non-zero only for $k=-k'$. The conjugate pairs are therefore $(\eta_+,\phi_-)$ and $(\eta_-,\phi_+)$. Hamilton's
equations become
\begin{equation}\label{eq:split_eom}
    \partial_t\eta_+=\frac{\delta\mathcal{H}^{\mathrm{phys}}}{\delta \phi_-},\quad
    \partial_t\phi_+=-\frac{\delta\mathcal{H}^{\mathrm{phys}}}{\delta \eta_-},\quad
    \partial_t\eta_-=\frac{\delta\mathcal{H}^{\mathrm{phys}}}{\delta \phi_+},\quad
    \partial_t\phi_-=-\frac{\delta\mathcal{H}^{\mathrm{phys}}}{\delta \eta_+}.
\end{equation}
If we now consider only one-sided data,
\begin{equation}\label{one_sided_data}
    \eta=\eta_+,\qquad \phi=\phi_+,\qquad\eta_-=\phi_-=0,
\end{equation}
applied in \eqref{eq:split_eom} after the variational derivatives are calculated, we obtain a
tool that reads the Hamiltonian selectively. A term of $\mathcal{H}^{\mathrm{phys}}$ now
contributes to $\partial_t\eta_+$ or $\partial_t\phi_+$ only if it contains exactly one factor
$\eta_-$ or $\phi_-$, which the variational derivative removes, and no second one, which would
make the term vanish on the one-sided data. The right-hand sides therefore collect exactly the
terms with a single negative wavenumber, in other words only the contributions from the
wavenumber-sign sectors 1-1, 2-1, 3-1, \dots of the Hamiltonian, and nothing else.
The equations for $\eta_-$ and $\phi_-$ would need a term containing only $\eta_+$ and $\phi_+$, which momentum conservation excludes, so one-sided data stay one-sided. Note that such
data do not correspond to any physical state, as a real field has content at $\pm k$; they are
used here as a formal device for reading only the $(n{-}1)$-$1$ sectors of the Hamiltonian.

\subsection{Exact problem on one-sided data}

A further reason for using the restriction \eqref{one_sided_data} is that the exact Euler
equations can then be written in closed form on the surface, and they are exactly the
first two equations of \eqref{eq:split_eom} evaluated on one-sided data. The
variational derivatives $\delta\mathcal H^{\mathrm{phys}}/\delta\phi_-$ and
$\delta\mathcal H^{\mathrm{phys}}/\delta\eta_-$
on such data are thus obtained
at all orders without using the Hamiltonian at all.

The general solution of the Laplace equation $\Phi_{xx}+\Phi_{yy}=0$ in deep water is
\begin{multline}\label{eq:Phi_split}
    \Phi(x,y)=\frac{1}{\sqrt{2\pi}}\int B(k)\mathrm{e}^{\ii kx + |k|y}\,\mathrm{d}k\\=\frac{1}{\sqrt{2\pi}}\int_0^\infty\left[ B(k)\mathrm{e}^{\ii k(x-\ii y)}+B(-k)\mathrm{e}^{-\ii k(x+\ii y)}\right]\mathrm{d}k\\=\Psi_+(x-\ii y)+\Psi_-(x+\ii y),
\end{multline}
where $B(k)$ is the Fourier transform of the trace on the mean level,
$\Phi(x,0)=(2\pi)^{-1/2}\!\int B(k)\,\mathrm{e}^{\ii kx}\,\mathrm{d}k$, and
\begin{equation}
\Psi_+(z)=\frac{1}{\sqrt{2\pi}}\int_0^\infty B(k)\,\mathrm{e}^{\ii kz}\,\mathrm{d}k,\qquad
\Psi_-(z)=\frac{1}{\sqrt{2\pi}}\int_0^\infty B(-k)\,\mathrm{e}^{-\ii kz}\,\mathrm{d}k
\end{equation}
are analytic in $\operatorname{Im} z>0$ and $\operatorname{Im} z<0$ respectively,
so that both terms are analytic throughout the fluid $y<0$. For
real $\Phi$, $B(-k)=\overline{B(k)}$, hence $\Psi_-(z)=\overline{\Psi_+(\bar z)}$
and $\Phi=2\,\mathrm{Re}\,\Psi_+(x-\ii y)$.

We see that the deep-water problem in one horizontal dimension splits naturally: as a single-variable function, $\Psi_+(x-\ii y)$ has only positive-wavenumber content, and $\Psi_-(x+\ii y)$ only negative-wavenumber content.
The dynamic pressure term of the Bernoulli equation becomes
\begin{equation}\label{dyn_pressure_complex}
    \frac{1}{2}\left[(\Phi_x)^2+(\Phi_y)^2\right]=2\Psi'_+(x-\ii y)\Psi'_-(x+\ii y),
\end{equation}
where prime denotes an ordinary derivative. Equation \eqref{dyn_pressure_complex} is a classical result related to the complex velocity potential \citep{lamb1932hydrodynamics}. Here we use the fact that it is the product of positive- and negative-wavenumber halves, with no other terms, and is therefore zero on one-sided data.

With \eqref{one_sided_data} and $\Psi_-\equiv0$, the surface value of the velocity potential has the form
\begin{equation}\label{phi_Psi_pluss}
    \phi(x,t)=\Phi(x,y,t)|_{y=\epsilon\eta}=\Psi_+(x-\ii\epsilon\eta(x,t),t),
\end{equation}
and it determines $\Phi=\Psi_+(x-\ii y, t)$ throughout the fluid since the condition can be solved for $\Psi_+$ order by order in $\epsilon$ within functions of positive-wavenumber content.
Differentiating $\phi$ with respect to $x$ and $t$, and using
$
    \Phi_y=-\ii\,\Psi_+'(x-\ii y)=-\ii\, \Phi_x,
$
we have for the surface values of the derivatives
\begin{align}\label{surf_1}
    \Phi_x|_{y=\epsilon\eta}&\,=\,\ii\,\Phi_y|_{y=\epsilon\eta}\,=\,\frac{\phi_x}{1-\ii\epsilon\eta_x},\\\label{surf_2}
    \Phi_t|_{y=\epsilon\eta}&\,=\,\phi_t-\epsilon\eta_t\Phi_y|_{y=\epsilon\eta}\,=\,\phi_t+\frac{\ii\epsilon\eta_t\phi_x}{1-\ii\epsilon\eta_x}.
\end{align}
The exact kinematic and dynamic boundary conditions become a closed system
\begin{align}\label{onesided_kinematic}
    \eta_t+\left[\epsilon\eta_x\Phi_x-\Phi_y\right]_{y=\epsilon\eta}&\,=\,\eta_t+\ii\phi_x=0,\\\label{onesided_dynamic}
    \eta+\left[\Phi_t+\frac{\epsilon}{2}\left(\Phi_x^2+\Phi_y^2\right)\right]_{y=\epsilon\eta}&\,=\,\eta+\phi_t+\frac{\ii\epsilon\eta_t\phi_x}{1-\ii\epsilon\eta_x}=0.
\end{align}
Eliminating $\eta_t$ from \eqref{onesided_dynamic} gives the exact evolution equations on one-sided fields
\begin{equation}\label{euler_onesided}
\eta_t=-\ii\phi_x,\qquad
\phi_t=-\eta-\frac{\epsilon\,\phi_x^2}{1-\ii\epsilon\eta_x}.
\end{equation}
From the first equation we see that the Dirichlet-to-Neumann operator $G(\eta)$ \citep{CraigSulem1993} has reduced to its flat-surface form $G(0)=|\partial_x|$, which on one-sided data is $-\ii\partial_x$. The only surviving nonlinearity, in the second equation, comes from the surface-motion term
$\epsilon\eta_t\Phi_y|_{y=\epsilon\eta}$ of \eqref{surf_2}, which accounts for
the change in the surface potential as the free surface moves vertically. It is
purely kinematic, a change of the point at which the potential is evaluated, and
the exact flow of $F$ removes this term, as shown in the following subsections.

The system \eqref{euler_onesided} was obtained in \citet{Ussembayev2019b}
by summing the kernel series generated by the recursion of
\citet{Ussembayev2019}, and was interpreted there as the exact equations
for unidirectional waves. That reading needs qualification. A real
unidirectional wave field has Fourier content at both signs of the
wavenumber and does not satisfy \eqref{euler_onesided}. Moreover, the
system carries no nonlinear dynamics, as shown below, and it is precisely this
absence that at quartic order produces the vanishing of the non-trivial resonant four-wave
interactions.

\subsection{Exact flow of the cubic generator}\label{exact_flow_section}

As already mentioned in \S\ref{sec:setup}, the transformation eliminating the cubic terms can be defined as the Hamiltonian flow of the functional
\begin{equation}
    F[u,v]=-\tfrac{1}{2}\int(Hu)^2Hv_x\,\mathrm{d}x,
\end{equation}
which satisfies $\{\mathcal H_2^{\mathrm{phys}},F\}=-\mathcal H_3^{\mathrm{phys}}$ \citep[see for example][and references therein]{CraigSulem2016,CraigGuyenneSulem2021, GuyenneKairzhanSulem2022}.
The flow is determined by Hamilton's equations
\begin{equation}\label{eq:flowF}
\tilde u_s=\frac{\delta F}{\delta v}\Big|_{(\tilde u,\tilde v)}=-H(H\tilde u\,H\tilde u_x),\qquad \tilde v_s=-\frac{\delta F}{\delta u}\Big|_{(\tilde u,\tilde v)}=-H(H\tilde u\,H\tilde v_x),
\end{equation}
where $(\tilde u, \tilde v)$ denotes the moving point of the flow and $s$ is the flow parameter.
At $s=0$ we set $\tilde u = u$, $\tilde v = v$, and at $s=\epsilon$ we have $\tilde u = \eta$, $\tilde v = \phi$, recovering the physical variables. This defines the exact transformation to all orders in $\epsilon$. By defining the variables
\begin{equation}
    \rho = H\tilde u, \qquad \mu = H\tilde v, 
\end{equation}
the flow equations \eqref{eq:flowF} become \citep{CraigSulem2016}
\begin{equation}
    \rho_s=\rho\rho_x, \qquad \mu_s=\rho\mu_x.
\end{equation}
The first one is an inviscid Burgers' equation \citep{Whitham, debnath2005nonlinear}, and the second one is a linear transport equation. The general solution of the system is
\begin{equation}
    \rho(s,x)=f_1(x+\rho(s,x) s), \qquad  \mu(s,x)=f_2(x+\rho(s,x) s),
\end{equation}
where $f_1$ and $f_2$ are arbitrary functions. At $s=0$ we have 
\begin{equation}
    \rho(0,x)=f_1(x)=(Hu)(x),\qquad \mu(0,x)=f_2(x)=(Hv)(x),
\end{equation}
determining the arbitrary functions, and at $s=\epsilon$ we get
\begin{align}
    \rho(\epsilon, x)&=(Hu)(x+\rho(\epsilon, x)\epsilon)=(H\eta)(x),\\ \mu(\epsilon, x)&=(Hv)(x+\rho(\epsilon, x)\epsilon)=(H\phi)(x).
\end{align}
Hence the exact $(u,v)\mapsto(\eta,\phi)$ transformation (the time-$\epsilon$ map of the flow) generated by $F$ is
\begin{equation}\label{full_flow}
    \eta = -H[(Hu)(x+\epsilon H\eta)],\qquad
    \phi = -H[(Hv)(x+\epsilon H\eta)].
\end{equation}
These implicit formulas hold as long as the characteristics of the Burgers'
equation above do not cross, that is, for $\epsilon\max_x(Hu)_x<1$, which is the
case throughout the weakly nonlinear regime.
Expanding through $\mathcal{O}(\epsilon^2)$ reproduces \eqref{eq:can_eta}--\eqref{eq:can_phi}.

\subsection{Selection rule for the full Hamiltonian}

On positive one-sided data, $u=u_+$, $v=v_+$, the Hilbert transform $H$ acts as multiplication by $-\ii$ and the exact transformation \eqref{full_flow} reduces to
\begin{equation}\label{eq:shift}
\eta(x,t)=u\big(x-\ii\epsilon\eta(x,t),t\big),\qquad
\phi(x,t)=v\big(x-\ii\epsilon\eta(x,t),t\big).
\end{equation}
With \eqref{phi_Psi_pluss} we see that  here $v(x,t)=\Psi_+(x,t)$ is the potential on the fixed mean level $y=0$, whereas $\phi(x,t)$ is the potential on the
moving surface $y=\epsilon\eta(x,t)$.
Differentiating these expressions and inserting into \eqref{euler_onesided} gives, after simplification, a linear system for one-sided fields:
\begin{equation}\label{one-sided-F-frame-equations}
    u_t=-\ii v_x,\qquad v_t=-u.
\end{equation} 
This is the main result of this section. We see that in the variables of the cubic flow, the exact water-wave problem on one-sided data, with the
full kinematic and dynamic boundary conditions and no truncation in $\epsilon$, is the linear problem.
By Hamilton's equations in the form \eqref{eq:split_eom}, written now for $u=u_++u_-$ and $v=v_++v_-$,
this is equivalent to
\begin{equation}\label{hamiltonian_evaluated}
    \left.\frac{\delta \mathcal{H}}{\delta u_-}\right|_{u_-=v_-=0}=u_+, \qquad \left.\frac{\delta \mathcal{H}}{\delta v_-}\right|_{u_-=v_-=0}=|\partial_x|v_+,
\end{equation}
where $\mathcal{H}[u_+,v_+,u_-,v_-]$ is the full Hamiltonian in these variables.
The right-hand sides are exactly those of the quadratic Hamiltonian,
\begin{equation}\label{H2_onesided_uv}
    \mathcal{H}_2=\frac{1}{2}\int (u^2+v|\partial_x|v)\,\mathrm{d}x=\int (u_- u_+ + v_-|\partial_x|v_+)\,\mathrm{d}x,
\end{equation}
so the terms of $\mathcal{H}$ beyond $\mathcal{H}_2$ contribute nothing to
\eqref{hamiltonian_evaluated}. The same argument repeated for negative one-sided data, $u=u_-$, $v=v_-$, gives \eqref{hamiltonian_evaluated} with all the indices switched.

By the reading of \S\ref{sec:decomposition}, the terms of $\mathcal{H}$ that contribute to
\eqref{hamiltonian_evaluated} are those with exactly one factor $u_-$ or $v_-$, as in \eqref{H2_onesided_uv}. Such a term
of $\mathcal{H}_n$ would add to the right-hand sides an expression of degree $n-1$ in $u_+$
and $v_+$, and the right-hand sides are linear. Hence there is no such term for $n>2$,
and likewise with the index signs interchanged. This is the \emph{selection rule}:
\begin{quote}\normalsize\itshape
In the frame of the exact flow of $F$, no term of the Hamiltonian beyond the quadratic part
has a single wavenumber of the minority sign, at any order, on or off the resonant manifold.
\end{quote}
Since momentum conservation excludes terms with wavenumbers of one sign only, every term beyond the quadratic part
then involves at least two positive and at least two negative wavenumbers. All $n$-0 and $(n{-}1)$-$1$ wavenumber-sign sectors of $\mathcal{H}_n$ at every order $n>2$ are empty.

The non-aligned 2-2 zero of \S\ref{sec:kernel_and_signs} lies outside of this selection rule. It
concerns terms with exactly one $u_-$ ($u_+$) and one $v_-$ ($v_+$) factor.
Our numerical computations show that it extends to order $n=5$, suggesting a general
pattern for $(n{-}2)$-$2$ wavenumber-sign sectors: the kernel of $\mathcal{H}_n$, with $n>2$, vanishes whenever the wavenumbers of one sign are a single
$u$-wavenumber and a single $v$-wavenumber. We do not have a proof of this and do not pursue it in this
paper.

\subsection{Structure order by order}\label{sec:order_by_order}

Three wavenumbers cannot include two of each sign, so at $n=3$ the selection rule reads
$\mathcal{H}_3=0$, the defining property of the $(u,v)$ frame.

For $n=4$ the rule is the vanishing of the 3-1 sector, which by \S\ref{sec:sectors} is the invariance of
$\mathcal{H}_4$ under the Hilbert rotation \eqref{eq:hilbrot}. 
Note that the analytic signals of \S\ref{sec:symmetry_and_W} are $p=2u_+$ and $q=2|\partial_x|v_+$, and the Hilbert rotation \eqref{eq:hilbrot} multiplies $u_+$, $v_+$ by $\mathrm{e}^{-\ii\alpha}$ and $u_-$, $v_-$ by $\mathrm{e}^{\ii\alpha}$.

The instances $n=3$ and $n=4$ also explain two facts noted in \citet{Simson2026}: a monochromatic
wave is an exact solution of the truncated equations (\S\ref{sec:setup}), and the envelope of a
narrow-band wave train has no second or third harmonic in these variables. Since the canonical bracket pairs opposite wavenumbers (\S\ref{sec:decomposition}), a Fourier
component with wavenumber $k$ is driven by the terms of the Hamiltonian that contain the
wavenumber $-k$, each through the product of its other factors. Components with wavenumbers
$k_a$, $k_b$, $k_c$ of one sign could therefore force the components $k_a+k_b$ and
$k_a+k_b+k_c$ only through terms with the wavenumbers $(k_a,k_b,-k_a-k_b)$ and
$(k_a,k_b,k_c,-k_a-k_b-k_c)$, which belong to the empty 2-1 and 3-1 sectors. The bound
harmonics of the Stokes wave and of the wave train are therefore carried by the transformation
\eqref{eq:can_eta} alone.

At $n=5$ the rule empties the 4-1 sector, of charge $\pm3$, on and off the resonant manifold,
and leaves only the 3-2 sector, of charge $\pm1$. From this order onwards, the frame is that of the exact
flow \eqref{full_flow}, since $\mathcal{H}_5$ depends on the $O(\epsilon^3)$ term of the transformation, which
\eqref{eq:can_eta}--\eqref{eq:can_phi} do not contain.

To compare with the five-wave coefficients of the literature, we pass to the normal variables
\eqref{normal_variables}. Every resonant quintet is known to be of the $3\leftrightarrow2$ interaction class only, $(k_1,k_2,k_3)\to (k_4,k_5)$
\citep{DyachenkoLvovZakharov1995,LVOV199738}. As in \S\ref{sec:zakharov_section}, the outgoing wavenumbers enter the real kernel
with reversed sign, so the sector of the quintet is that of $(k_1,k_2,k_3,-k_4,-k_5)$:
3-2 or 4-1, depending on the signs of the wavenumbers. 
Its coefficient in the literature is that of the reduced Zakharov Hamiltonian, and it is the
same as the coefficient of the $3\leftrightarrow2$ class of $\mathcal{H}_5[u,v]$. Indeed, the reduced
Hamiltonian is reached from the present frame by the quartic and quintic steps of the
reduction. A quartic generator $F_4$ acts on $\mathcal{H}_n$ through Poisson brackets, the
first of which, $\{\mathcal{H}_n,F_4\}$, is of degree $n+2$. A quintic term can therefore arise
only from $\{\mathcal{H}_3,F_4\}$, and $\mathcal{H}_3=0$, so the quartic step leaves
$\mathcal{H}_5$ unchanged. The quintic step removes the other classes of $\mathcal{H}_5$,
$4\leftrightarrow1$ and $5\leftrightarrow0$, which are never resonant, and, taken without a
$3\leftrightarrow2$ part of its own, leaves this class as it is. 
As at fourth order (\S\ref{sec:vanishing_demonstration}), its value on
the resonant manifold does not depend on the variables. By the selection rule, a resonant quintet of the 4-1 sector
therefore has zero coefficient in any variables, while the rule says nothing about the 3-2
sector.

The five-wave coefficients computed in the literature
follow this pattern. The co-propagating quintets, all five waves moving in
the same direction, lie in the 3-2 sector. Their coefficient is non-zero
\citep[eq.~5.7]{DyachenkoLvovZakharov1995}, which is the proof of
non-integrability at fifth order. \citet{LVOV199738} computed the
coefficient for every arrangement of signs among the five wavenumbers, five
cases up to symmetry, and found it non-zero in three of them and zero in two,
adding that the reason for the zeros remained to be explained. The three cases form the 3-2
sector and the two form the 4-1 sector. The zero found earlier by
\citet[\S2.4]{BryantStiassnie1994} in the standing-wave problem, for the
quintet $(k,k,k)\to(4k,-k)$ in which three co-propagating waves convert into
a right- and a left-moving one, is a 4-1 zero as well.

Physically, at fifth order a unidirectional sea is no longer closed, but it
opens in one way only: a pair of right-moving waves may emit, or absorb, a
single left-moving wave, $2R\leftrightarrow 2R+1L$, with the coefficients computed by \citet{LVOV199738}, whereas the conversion of three right-moving waves
into a right- and a left-moving one, $3R\to 1R+1L$, is absent. In particular, a single Stokes wave of wavenumber $k$ (expressed by three identical $k$ values in the incoming slots) does not resonantly generate
a free Stokes wave at $4k$ together with a left-moving Stokes wave at $-k$. This would be the $(k,k,k)\to(4k,-k)$ process from above, and momentum
and energy conservation allow it, as $3k = 4k - k$ and
$3\sqrt{k} = \sqrt{4k} + \sqrt{k}$. It is forbidden by the selection rule. In terms of $W=N^+-N^-$, this conversion would change the
imbalance by three units, the charge of the empty 4-1 sector, whereas the emission of a single
left-moving wave changes it by one, the charge of the 3-2 sector. The imbalance is not conserved beyond fourth order, but it changes by exactly one unit in every
resonant quintet, and by at most $n-4$ units at order $n$, the largest charge that the
selection rule leaves (\S\ref{sec:sectors}).

\subsection{What the symmetry and the selection rule depend on}

The following assumptions were used to obtain the selection rule: the fluid has no bottom, the motion has one
horizontal dimension, and there is no restoring force other than gravity. The first two give the splitting
\eqref{eq:Phi_split}, so that the quadratic term of Bernoulli's equation is the product \eqref{dyn_pressure_complex}, which
vanishes on one-sided data, and the kinematic boundary condition is linear in that case. The third leaves only the term $\eta$ in the dynamic boundary condition, and the shift \eqref{eq:shift} carries it unchanged.
If any one of the three assumptions is dropped, the selection rule fails, and with it the symmetry, as we now discuss.

In water of depth $h$ the harmonic extension of $\mathrm{e}^{\ii kx}$ is
$\mathrm{e}^{\ii kx}\cosh k(y+h)/\cosh kh$ \citep{lamb1932hydrodynamics}. It consists of a part
in $x-\ii y$ and a reflected part in $x+\ii y$, of relative weight $\mathrm{e}^{-2kh}$, and for
a single mode $\Phi_x^2+\Phi_y^2=-k^2\,\mathrm{e}^{2\ii kx}/\cosh^2 kh$. The two parts thus no longer carry one sign of the wavenumber each, so their product, the
pressure term, does not vanish on one-sided data, and the derivation above does not go through.

With two horizontal dimensions the harmonic extension of
$\mathrm{e}^{\ii\boldsymbol{k}\cdot\boldsymbol{x}}$ is
$\mathrm{e}^{\ii\boldsymbol{k}\cdot\boldsymbol{x}+|\boldsymbol{k}|y}$, and two such modes
contribute to $\tfrac12\nabla\Phi\cdot\nabla\Phi$ a term proportional to
$|\boldsymbol{k}||\boldsymbol{k}'|(1-\cos\theta)$, where $\theta$ is the angle between the
wavevectors. It vanishes only when the two wavevectors have the same direction, that is, only for the
one-sided data of the one-dimensional problem. With two horizontal dimensions the non-trivial
resonant quartets are known to interact
\citep{Phillips1960,LonguetHiggins1962,McGoldrickEtAl1966}, and they carry the energy transfer
described by the kinetic equation \citep{Hasselmann1962}.

The role of gravity (and of the dispersion relation) is seen in the generator $F$ (\S\ref{exact_flow_section}). The shift
\eqref{eq:shift} is the flow of $F$ on one-sided
data and does not depend on the potential energy. What depends on it is whether the shift makes
the one-sided problem linear, and for that, $F$ must at least remove all the cubic terms. For a
general quadratic potential energy $\mathcal{U}=\tfrac12\int\eta\,\Omega\,\eta\,\mathrm{d}x$
with symbol $\Omega(k)$, the transformation generated by $F$ produces the cubic term
\begin{equation}
    \{\mathcal{U},F\}=-\tfrac12\int(H\eta)^2\,\Omega H\eta_x\,\mathrm{d}x,
\end{equation}
whose kernel is proportional to $\omega^2(k_1+k_2)-\omega^2(k_1)-\omega^2(k_2)$, with
$\omega^2=|k|\,\Omega(k)$ and $k_1,k_2$ of one sign. It vanishes identically only for $\omega^2\propto|k|$. For gravity, $\Omega=1$, the integrand
is the total derivative $-\tfrac16\partial_x(H\eta)^3$, so the potential energy is left
unchanged and the equation that $F$ satisfies involves the kinetic energy only. $F$ then
contains no frequencies, which is why the kernel \eqref{eq:K-closed} has no frequency
denominators (\S\ref{sec:kernel}). Moreover, $\omega=\sqrt{|k|}$ is concave and admits no
three-wave resonances, so the generator $F$ is unique and the frame in which the selection rule
holds is fixed. For any other dispersion relation the cubic terms, where they can be removed,
are removed by a different generator. Its flow is not the shift \eqref{eq:shift}, and the argument that led to
the selection rule does not apply to its frame. For surface tension, $\Omega=1+\gamma k^2$, the
cubic term produced by $F$ is $\tfrac12\gamma\int(H\eta_x)^3\,\mathrm{d}x$, whose kernel,
proportional to $\gamma|k_1k_2(k_1+k_2)|$, vanishes nowhere, so already the linear part of the
surface tension prevents $F$ from removing the cubic terms. At the three-wave resonances of
gravity-capillary waves \citep{McGoldrick1965,McGoldrick1970} they cannot be removed at all, and
a cubic term always has a single wavenumber of the minority sign
(\S\ref{sec:order_by_order}), so the selection rule fails already at third order, in any
variables.

Physically, we can thus connect the symmetry and the selection rule with the dispersion relation
$\omega^2=|k|$ of planar deep-water gravity waves. The property of $\omega^2$ that the
derivation requires is additivity, $\omega^2(k_1+k_2)=\omega^2(k_1)+\omega^2(k_2)$ for
wavenumbers of one sign. It is what makes the pressure term vanish on one-sided data, since in deep water the kernel of
the pressure term is proportional to $(|k_1|+|k_2|)^2-|k_1+k_2|^2$. And it is what leaves the
potential energy unchanged by the shift, since the cubic term that $F$ produces from it has a
kernel proportional to $\omega^2(k_1+k_2)-\omega^2(k_1)-\omega^2(k_2)$. For $\omega^2=|k|$ both vanish identically when $k_1$ and $k_2$ have the same sign. Each of the standard
extensions destroys this property. At finite depth $\omega^2=|k|\tanh|k|h$, with surface tension
$\omega^2=|k|+\gamma|k|^3$, and with two horizontal dimensions
$|\boldsymbol{k}_1+\boldsymbol{k}_2|<|\boldsymbol{k}_1|+|\boldsymbol{k}_2|$ unless the
wavevectors have the same direction.

\section{Conclusions and discussion}

We showed that in the canonical variables that remove the cubic terms,
the frame of the flow of the cubic generator, the deep-water Hamiltonian
through quartic order is invariant under the Hilbert rotation
\eqref{eq:hilbrot}. Its Noether charge $W$
is the integral anticipated by \citet{DyachenkoZakharov1994}. The
invariance is the statement that the quartic kernel is supported only on
the 2-2 wavenumber-sign sector. The non-trivial resonant quartets lie in
the 3-1 sector, so their coefficients vanish without calculation. In
complex normal variables, the $2\leftrightarrow2$ interaction class of the
real kernel \eqref{eq:K-momentum} is the closed form of the reduced
Zakharov kernel \citep{DyachenkoKachulinZakharov2017,Dyachenko2020canonical},
off the manifold included, with DKZ's sign rule as the 3-1 zero.

The symmetry is the quartic instance of a
selection rule that holds at every order in the frame of the exact flow
of the cubic generator: no term of the Hamiltonian has a single
wavenumber carrying the minority sign; the quintic instance is the five-wave
zeros of \citet{LVOV199738}. Its origin is a property of planar deep-water gravity waves. The harmonic field splits into two holomorphic functions of
$x\mp\ii y$, each carrying one sign of the wavenumbers, and in the frame
of the cubic flow the exact problem is linear on one-sided data. This
linearity is the selection rule (\S\ref{sec:origin}).
Finite depth, a second horizontal dimension and surface tension each break
the rule, and the quintic 3-2 sector breaks the Hilbert rotation
symmetry, which holds only for the quartic truncation.

The reduced Zakharov formulation hides this structure. The complex normal variables merge the two real fields
and the two signs of the wavenumber, and the information about wavenumber
signs is spread over many monomials. The reduction eliminates the cubic terms and the number-changing quartic
classes, neither of which carries long-time dynamics. The quartic term
that remains comes out as the quartic kernel of the original Hamiltonian plus six exchange terms with
frequency denominators, none of which vanishes individually on the 3-1
wavenumber-sign sector. The value of the four-wave coefficient on the resonant
manifold is the one thing no near-identity canonical transformation can
change. In the frame of the cubic flow the whole sector is empty (\S\ref{sec:origin}); in every other frame (free of cubic terms) the zero
must still appear on the manifold, and in the reduced Zakharov formulation it appears without its reason, as
a sum that happens to vanish identically.

The symmetry is a property of the kernel on the whole momentum
shell, and the standard theories usually examine only parts of the
shell: the kinetic equation uses only the resonant manifold, envelope
theory only the neighbourhood of the carrier wavenumber, and
unidirectional theory only the co-propagating sector closed by DKZ's
sign rule. It needed the whole shell to be seen. The frame in which the reason
is visible is reached by the cubic elimination alone. It is usually only
an intermediate frame, passed through on the way to the reduced
equations, and the procedure itself gives no reason to pause there.

To our knowledge, the Hilbert rotation \eqref{eq:hilbrot} has not
previously been stated as a symmetry of a truncated water-wave
Hamiltonian, although many of its consequences, as listed in
\S\ref{sec:intro} and throughout the paper, are in print.
\citet{DyachenkoZakharov1994} wrote that the cancellation ``apparently is
not accidental''. We showed that it is the quartic shadow of a rule that holds
at every order and rests on the holomorphic structure of planar
deep-water flow and the square-root dispersion relation of gravity waves.

\begin{bmhead}[Acknowledgements.]
The author thanks the anonymous referees of \citet{Simson2026}, whose comments prompted this work: one drew attention to \citet{Dyachenko2020canonical}, and another observed that the canonical transformation used there is the Taylor expansion of the Hamiltonian flow of \citet{CraigGuyenneSulem2021}.
Helpful discussions with Jaan Kalda, Dmitri Kartofelev, and Andrus Salupere are also gratefully acknowledged.
\end{bmhead}

\begin{bmhead}[Declaration of AI use.]
During the preparation of this manuscript (August--November 2026) the author used generative AI tools (Claude and Gemini) for language editing, for revising some parts of the text, for checking the author's derivations, and for writing Python scripts that verify them numerically. All AI-assisted text was revised by the author, who takes full responsibility for the final content. The scientific ideas and results of this paper are the author's.

\end{bmhead}

\begin{bmhead}[Funding.]
This research received no specific grant from any funding agency,
commercial or not-for-profit sectors.
\end{bmhead}

\begin{bmhead}[Declaration of interests.]
The author reports no conflicts of interest.
\end{bmhead}

\bibliographystyle{jfm}
\bibliography{paivos_bib}

\end{document}